\documentclass[doublecol,figures]{epl2}

\title{Unjamming a granular hopper by vibration}

\author{A. Janda\inst{1} \and D. Maza\inst{1} \and A. Garcimart\'{\i}n\inst{1} \and E. Kolb\inst{2} \and J. Lanuza\inst{2} \and E. Cl\'ement\inst{2}}
\shortauthor{A. Janda \etal}

\institute{
\inst{1}Departamento de F\'{\i}sica, Facultad de Ciencias,
Universidad de Navarra, E-31080 Pamplona, Spain.\\
\inst{2} Laboratoire de Physique et M\'ecanique des Milieux
H\'et\'erog\`enes , associ\'e au CNRS (UMR 7636) à l'ESPCI et aux
Universit\'es Paris 6 et Paris 7, 10 rue Vauquelin, 75005 Paris
France.
}


\abstract{
We present an experimental study of the outflow of a hopper continuously vibrated by a piezoelectric device. Outpouring of grains can be achieved for apertures much below the usual jamming limit observed for non vibrated hoppers. Granular flow persists down to the physical limit of one grain diameter, a limit reached for a finite vibration amplitude. For the smaller orifices, we observe an intermittent regime characterized by alternated periods of flow and blockage. Vibrations do not significantly modify the flow rates both in the continuous and the intermittent regime. The analysis of the statistical features of the flowing regime shows that the flow time significantly increases with the vibration amplitude. However, at low vibration amplitude and small orifice sizes, the jamming time distribution displays an anomalous statistics.}

\begin{document}

\maketitle

\section{Introduction}

Since the end of the Middle Ages sand clocks have been customarily used to measure time flowing (see for example an historical monography in \cite{Junger}). Nowadays, hoppers are widely used in the industry to store and deliver granular matter\cite{Brown}. In both situations one uses a striking property of granular matter as the flowing rate does not depend on the height of material in the container but rather, on the size of the orifice. This was rationalized in the 60's by Beverloo, who proposed a simple scaling law for the discharge rate essentially based on the free falling velocity at a scale comparable to the aperture \cite{nedderman,beverloo}. The scaling law includes an empirical modification to account for small orifices (sizes of the order of a few grain diameters). In this case, one observes blocking of the flow due to the presence of static vaults forming at the exit \cite{to}. This can be problematic for industrial processes since it severely limits the possibility to deliver small quantities of material in a reliable way. In recent years this issue has been studied in detail with the aim to characterize more thoroughly the jamming probability and the flow properties near this limit \cite{to,eric1,janda1,zuriguel1,zuriguel2}. One way to deal in practice with the problem of blockages is to add vibrations in order to restore the flow by breaking the arches \cite{IMechE}. Previous works have been devoted to the influence of vibrations in a hopper discharge \cite{Hunt,Wassgren,Veje,Chen,Pacheco}. In most cases, however, the main interest was the influence of vibrations on the flow rate. As the motivation was usually a practical situation, these studies often deal with strong vibrations and large orifices, well above the jamming limit. Here we focus on small orifices, which are prone to jamming.
Indeed, this question is not just a technical one, but relates to the fundamental properties of granular matter and can be put in framework of current interrogations on the jamming transition of soft matter \cite{liunagel}. In the present case, deep interrogations arise about the  capability for a granular packing to flow in a restrained geometry and on the true nature of the block/unblock mechanisms.
From a theoretical point of view this set of questions can be framed in the context of the rigidity transition driven by the existence of a marginal mechanical state , i.e., where the average number of contacts is just sufficient to ensure a solid structure to the packing (the isostatic limit \cite{wyart}). In this case, one observes a large density of so-called "`soft modes"' corresponding to  collective granular motion at low energy cost. At the outlet of a hopper the confining pressure vanishes, thus giving a situation where, if of any importance, anomalous critical behavior should be revealed (at least from a finite size scaling perspective). In this paper, we present a novel insight on the problem, with an experimental design of a hopper that allows a fine statistical study of granular flows in this limit. The continuous variation of two crucial parameters (diameter opening and vibration amplitude) allows to deal with the probability of arch formation and thus, to go well below the usual jamming limit. In this way, we are able to investigate the properties of the granular flow down to the physical limit of one grain aperture.

\section{Experimental setup}

\begin{figure}
\onefigure[width=.92\columnwidth]{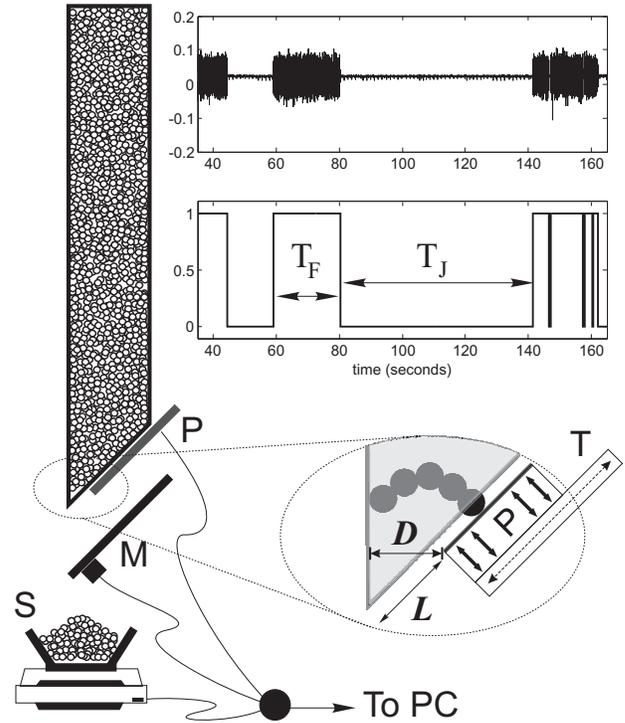}
\caption{Experimental set-up: M, microphone; S, scales; P, piezoelectric transducer; T, translation stage. The top of the granular column is open at the atmospheric pressure. A sketch of the bottom outlet is also shown with the variables $L$ and $D$. The piezoelectric transducer vibrates in the direction perpendicular to its surface as indicated by the black arrows. Note that the walls of the hopper are not touching the piezoelectric, but only the grains. The top right insets show the signal as measured from the microphone (\emph{top}, signal level in arbitrary units) and the binarized signal (\emph{bottom}).}
\label{setup}
\end{figure}

The silo is made of a plexiglass cylindrical tube (inner diameter 1.9 cm) and is specifically designed to control continuously two important parameters that will allow an extensive study of the transition to a blocking state, namely, the aperture opening and the vibration amplitude. To this purpose, the design of the bottom outlet is special. The cylinder's bottom edge was obtained by a planar cut at $45 \deg$. The resulting elliptic area is partially closed by a flat rectangular surface on which a thin piezoelectric transducer is mounted, its diameter being larger than the silo diameter. The blocking piece can be moved along the cut in the direction of the larger diameter of the ellipse via a micrometric translation stage (see Fig.~\ref{setup}). The hole hence produced, is characterized by a minimal opening distance $L$ (see inset of Fig.~\ref{setup}). The piezoelectric transducer is driven with a sinusoidal vibration at a frequency of 350 Hz and it vibrates in the direction perpendicular to its surface. Note that the vibration is directly applied to the granular material inside the silo, but not to the wall, which is not touching the vibrating transducer. The granular material consists of glass beads, their diameter being $1.55 \pm 0.05 ~ mm$. Prior to the experiments, we measured the energy input due to the vibrations induced by the piezoelectric device. An accelerometer was buried at the top of the fully filled silo and for different electric voltage amplitudes, the r.m.s. acceleration was monitored, with the outlet closed. A linear relation is obtained between the electric voltage and the acceleration in the bulk of the granular column. We checked that the acceleration does not change significantly when the outlet is open. The typical measured accelerations are only a fraction of the gravity acceleration. For the maximum induced vibration amplitude (a rms acceleration of about $1~m/s^2$), the typical kinetic energy is much smaller (by about five orders of magnitude) than the potential energy at the scale of a grain diameter. The mass flux is monitored by recording the mass of grains that fall from the silo on an electronic scale interfaced with a computer. To have a good determination of the time lapses when the orifice is jammed or when it flows, a plastic sheet with a microphone attached to it is mounted below the silo outlet. Thus the sound produced by the impact of the falling grains can be monitored and time lapses inferred.

\section{Dependence of the mean flow with the orifice size}

\begin{figure}
\onefigure[width=.92\columnwidth]{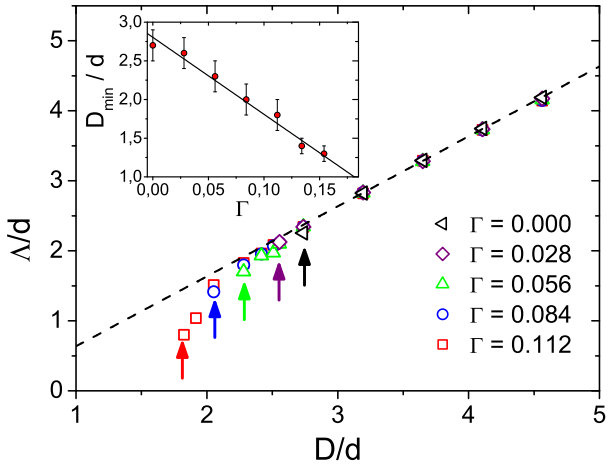}
\caption{(\textit{Color online}) Length scale $\Lambda$ normalized to the bead diameter $d$ as a function of rescaled aperture size $D$ for various vibration amplitudes. The dashed line is a fit of the data for big $D$; the slope is close to one. The arrows mark the minimum rescaled orifice size $D_{min}$ (definition explained in the text). \emph{Inset}: Value of $D_{min}/d$ as a function of $\Gamma$.}
\label{flow}
\end{figure}

The determination of the mean mass flow $Q$ was carried out by measuring the time taken to deliver a large quantity of grains (about $2\times10^{4}$). The conventional picture for the outpouring of grains through a hole is embodied in the so-called Beverloo's scaling \cite{beverloo,nedderman}, that relates the mean flow rate to the aperture diameter. In this scenario, grains are released from an \emph{imaginary} vault having the same radius as the exit orifice. In our case, it is not clear what is the right length scale $\Lambda$ controlling the size of the dynamical vault that actually releases the grains (and therefore determines the mean flow). So in order to benchmark our experimental setup, we  measured the mean flow $Q$ (in the absence of vibration) as a function of the size of the exit orifice for the cases where no permanent jams are observed. We have taken as a characteristic length the projection of $L$ (see Fig.~\ref{setup}) on the horizontal axis: $D=\frac{\sqrt{2}}{2}L$. To check whether the Beverloo scaling is valid, the length scale $\Lambda=(Qg^{-1/2}\rho^{-1})^{2/5}$, where $g$ is the gravity acceleration and $\rho$ the density of the beads, is computed. As seen in Fig.~\ref{flow}, the data without any vibration are compatible with the Beverloo scaling $Q\propto D^{5/2}$. Remark that we do not intend to discuss here the validity of this scaling, which is a priori not obvious for small orifices (and especially with the present geometry). This has been done elsewhere \cite{cristian}. The relevant point is that although the shape of the outlet is not circular and its area does not increase as $D^2$, the data follow conspicuously well Beverloo's scaling down to the blockage limit. This seems to indicate that there is only one structural length scale controlling the granular flux, and that it is directly linked to the aperture distance $D$.

\section{The influence of vibrations}

The applied vibration is quantified through the rescaled acceleration $\Gamma$, which is the rms acceleration normalized by the gravity acceleration $g$. Note that the pertinent parameter could also be the kinetic energy. We do not intend to discuss here the influence of the driving frequency, and only the driving amplitude at a given frequency is changed.

In the absence of vibration and for small apertures, there are situations where the flow is continuous for some time and then it can eventually get blocked forever. The application of a vibration can unjam the flow (it depends on how energetic the vibration is). With applied vibration, the flow can also get blocked, but the outpouring can resume after a finite time. In other words, an intermittent regime is observed. In order to extract rational flow properties we must define a strict and precise working protocol that involves the finite mass of grains we seek to deliver and the time we are ready to wait before deciding that the flow is halted. The protocol chosen involves a fixed amount of grains (approximately $2\times10^{4}$ beads); if this mass is delivered without any jam that lasts for $100~s$ or more, then the hopper is considered to be flowing. If there is a jam that lasts for $100~s$ or more, the run is finished and the silo is considered to be jammed. The time interval of $100~s$ was so chosen because our experiments indicate that it is very unlikely that the flow will resume by itself without vibration after that time lapse. For a given $\Gamma$ and $D$, the mean flow $Q$ is calculated as the mass discharged divided by the total time of measurement. The size of the orifice $D$ is then decreased and the measurements are carried out again, as long as the waiting time does not exceed the predefined limit. In this way, for each vibration amplitude we obtain $D_{min}(\Gamma)$, the smallest size of the orifice for which the flow has not been halted during a time lapse longer than $100~s$ before the mentioned amount of material is discharged (\emph{i. e.} the smallest value of $D$ for which the hopper is considered to be in a flowing state, as obtained with the mentioned protocol). In some way, $D_{min}$ might be related to the size of the orifice above which jamming is extremely improbable in a non vibrated silo \cite{zuriguel1}.

The data are displayed on Fig.~\ref{flow}, where the values of $D_{min}$ for different $\Gamma$ are indicated by arrows. The first salient feature is that for orifices of sizes above $D_{min}(\Gamma=0)$ (corresponding to the black arrow) the influence of vibration is negligible, i.e. the mass flux does not change whatever the value of $\Gamma$. Second, the vibrations allow the possibility of flow below this orifice size. The more energetic the vibration is, the lower is the minimal value of $D$ for which the silo can flow without getting jammed. The last striking point is that for a finite value of $\Gamma$, one can reach by linear extrapolation the value $D_{min}=1$ below which hard grains cannot go through the aperture (see inset of Fig.~\ref{flow}). This result proves conceptually the design value of such an activated outlet that is able to deliver granular flow down to the ultimate physical limit of one grain size ($D_{min}=1$). However, we need now to push the study further and characterize the flowing dynamics because below the value of $D_{min}(\Gamma=0)$ the mean flow as defined before does not follow anymore a simple Beverloo's scaling.

\section{Intermittency and Jamming}

\begin{figure}
\onefigure[width=.92\columnwidth]{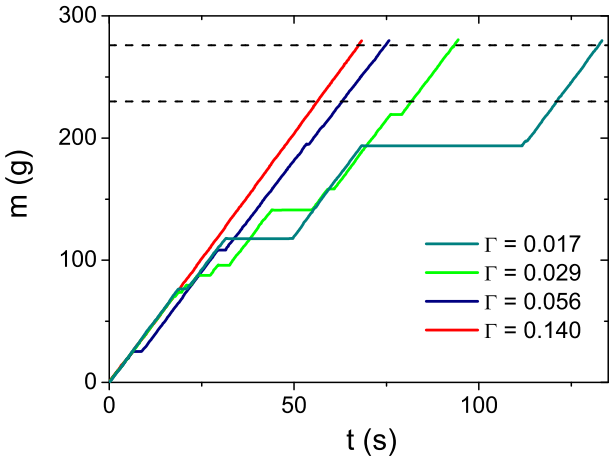}
\caption{(\textit{Color online}) The masses delivered by the silo with an exit orifice $D=2.42d$ as a function of time, for four different vibrations as indicated in the legend. The dashed horizontal lines mark the upper and lower bounds of the zone where the mean flow has been measured for comparison (see explanation in the text).}
\label{masse}
\end{figure}

In the presence of vibrations at the outlet, therefore, the usual limitation of a minimum size to deliver grains from a hopper can be overcome. However, in this regime the outlet flow displays a strong intermittent dynamics that needs to be characterized in more detail. Using the mass recorded on the scale, we examine directly the exit flow continuously over time. On Fig.~\ref{masse} four time series corresponding to the mass delivered on the scales are represented for four different vibration amplitudes, the exit orifice being the same --a certain value below $D_{min}(\Gamma=0)$. As the vibration decreases, the curves clearly display jam periods of larger duration. These events are rare or nonexistent for apertures larger than the value of $D_{min}(\Gamma=0)$.

From the data series as presented in the inset of Fig.~\ref{setup} (impacts of grains on the microphone) and in Fig.~\ref{masse} (outflow mass delivered on the scale), we propose that this dynamical issue can naturally be separated into two sets of questions pertaining to the flowing and to the jammed regimes: (i)when the hopper is flowing, how long is it going to flow and what is the flow rate? (ii)when it gets jammed, how long does the blockage persist and what fraction of the overall experimental time does it spend in that state?

These questions are going to be addressed in detail next.

\section{The flowing regime}

\begin{figure}
\onefigure[width=.92\columnwidth]{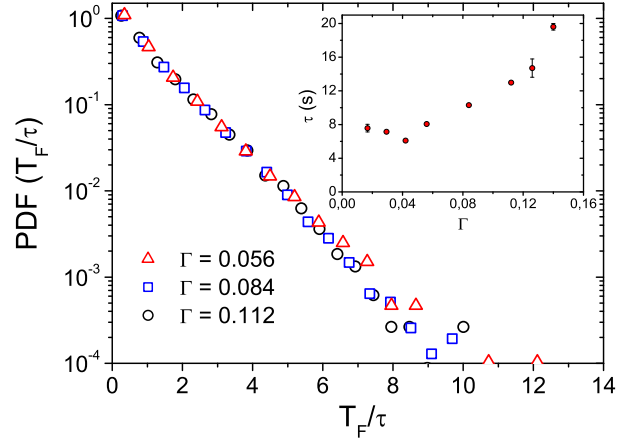}
\caption{(\textit{Color online}) Histogram of the rescaled time intervals during which the orifice $D=1.78d$ has been flowing  (in semilogarithmic scale). \emph{Inset}: Mean flowing time as a function of the rescaled rms acceleration $\Gamma$.}
\label{histtflow}
\end{figure}

\begin{figure}
\onefigure[width=.92\columnwidth]{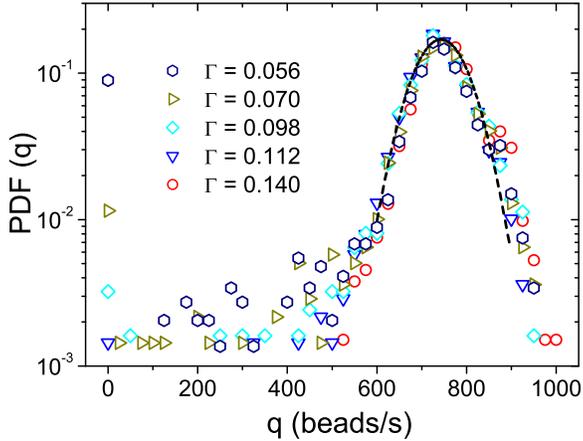}
\caption{(\textit{Color online}) PDFs of the flow rate (obtained by averaging in a time window of $1$ s) with an exit orifice $D=2.42d$ and for five different vibration amplitudes $\Gamma$ as indicated in the legend.}
\label{histoq}
\end{figure}

The time intervals $T_{F}$ during which the hopper is flowing are obtained from the impact of the out-flowing grains on the microphone (see Fig.~\ref{setup}, inset) for an experiment at constant $D$ and $\Gamma$. As seen in the Fig.~\ref{histtflow}, the distribution of $T_{F}$ shows an exponential tail. Therefore, a characteristic time $\tau(\Gamma,D)$ can be defined. The inset shows that this characteristic time increases with acceleration $\Gamma$ and goes to a finite value when $\Gamma$ tends to zero. The fact that one can define a characteristic value for the flowing intervals $T_{F}$ is consistent with the existence of mean avalanche sizes obtained in \cite{zuriguel1, zuriguel2} in the case of a non-vibrated silo where the discharge through the small orifice is restored by jet of pressurized air.  The remarkable point here is that this is also the case when vibrations are applied.
    This hints to an interesting feature that will be checked next, namely, that when the vibration is applied the restored flow has the same statistical features that the corresponding flow one would have without vibration. As noted earlier, this is true above $D_{min}(\Gamma=0)$. Let us take a closer look on Fig.~\ref{masse}, which corresponds to a diameter $D$ inside the intermittent domain, and in particular, let us consider four stretches in the plots where the orifice is not blocked (as marked with horizontal dashes). The mean flow for these intervals are all similar: 4.05, 3.99, 3.88 and $4.10 ~\pm 0.01$ g/s , respectively (the resolution is the least mean square fit error). To investigate this point in more detail, we computed the histograms of the flow rate $q$ corresponding to the same orifice size for five different vibration amplitudes (Fig.~\ref{histoq}). Note that in this case the flow $q$ is measured as the number of beads delivered in one second. The most remarkable feature of Fig.~\ref{histoq} is that the $q$ PDFs mainly consists of a Gaussian centered around the same mean value of the flow rate whatever $\Gamma$ is. The differences appear for the low flow rate events corresponding to the small flow rates. It seems that a higher vibration amplitude is able to reduce significantly those events that seems to bring the system in the vicinity of blockage. Consequently, the time of flow can be significantly increased as noted before. Interestingly, this distribution is in agreement with previous results for nonvibrated silos \cite{janda2}.

\section{The jammed regime}

\begin{figure}
\onefigure[width=.92\columnwidth]{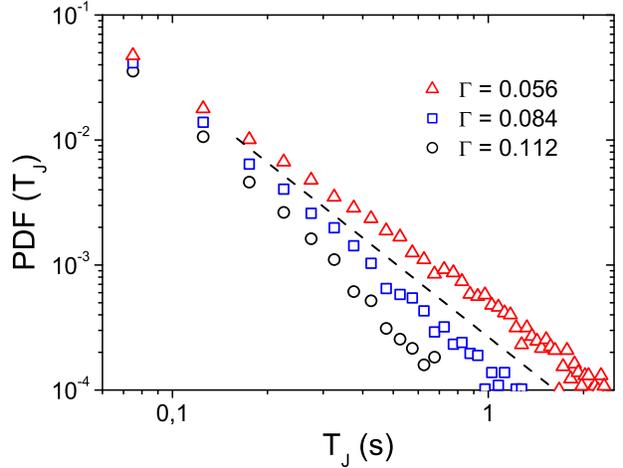}
\caption{(\textit{Color online}) Histogram for the time lapses that the orifice has remained blocked, in logarithmic scale, corresponding to $D=1.78d$. The dashed line has a slope of two. Note that for the smallest acceleration displayed (\emph{triangles}) the slope is smaller than two.}
\label{histtj}
\end{figure}

Let us now explore the jamming dynamics due to arch formation at the outlet of the hopper. These arches are constantly destroyed by vibrations and then recreated by jamming. The time intervals during which the outpouring has been halted are denoted by the variable $T_J$ (see inset of Fig.~\ref{setup}). The PDFs of $T_J$ for a fixed $D$ at several $\Gamma$ are displayed in Fig.~\ref{histtj}.

A typical time scale to compare with can be obtained by calculating the time it takes for a bead to fall its own diameter under the acceleration of gravity, namely $\tau_{g}=\sqrt{2 d /g} \approx 0.02 ~s$. For times larger than $\tau_{g}$ the distribution decays as a power law with an exponent that depends on the vibration amplitude (Fig.~\ref{histtj}). Importantly, for small holes and low $\Gamma$ the power law displays an exponent smaller than $2$ (absolute value), and this means that the first moment of the distribution does not converge. In other words, the data show that the distribution of jam durations can be extremely large and that the mean time during which the hopper is jammed will depend on the total time that the experiment has been running. Then the value of the mean flow can be dominated essentially by the longest jam. Due to these features, the term anomalous statistics is used here to describe the situation.

Since the transition to this regime can not be rigorously defined in terms of an average flow, we propose instead an alternative characterization. The fraction of time $\Phi$ for which the hopper is jammed is computed inside a time window of variable length ($T_{W}$). This time window is taken inside a long time series during which the experiment has been running. The measurements are carried out for a fixed diameter $D=2.42d$ and for several values of $\Gamma$.

\begin{figure}
\center{\includegraphics[width=.92\columnwidth]{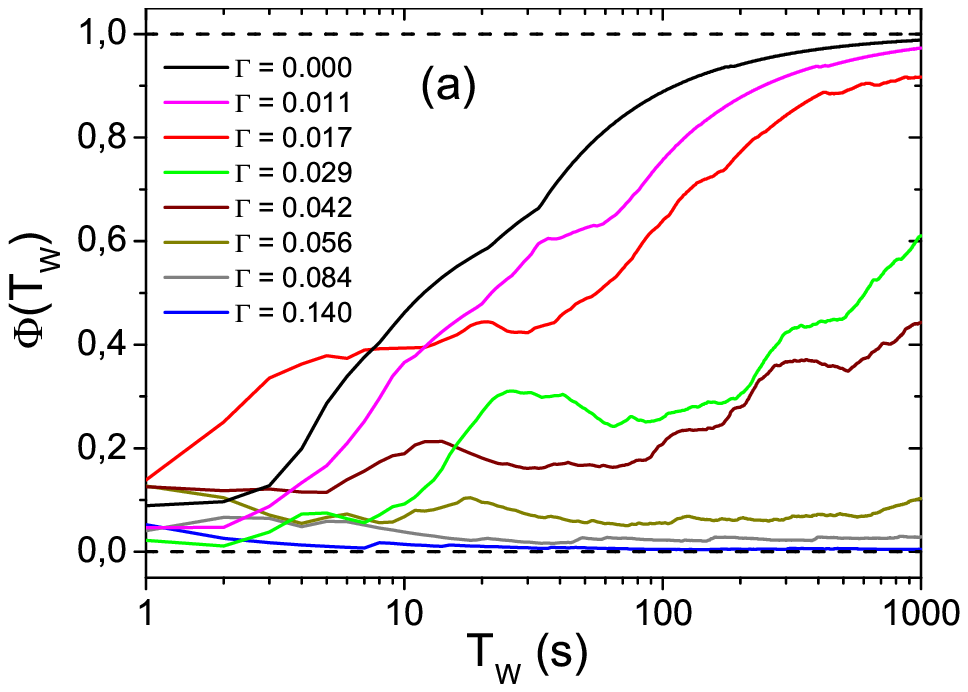}}
\center{\includegraphics[width=.92\columnwidth]{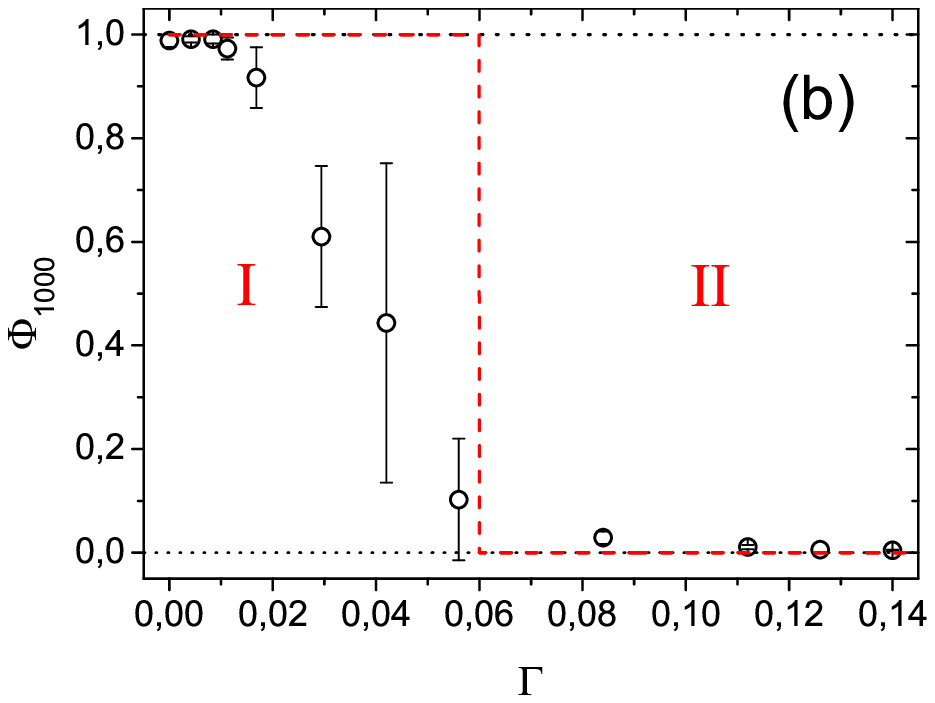}}
\caption{(\textit{Color online}) The transition from flowing to jammed regime. \textbf{(a)} Fraction of the time for which the flow is jammed $\Phi$ as a function of the observation time window $T_{W}$. Note the logarithmic scale on the horizontal axis. Each line is a single experimental run. \textbf{(b)} The fraction of time that the system is jammed, calculated for $T_{W}=1000s$, as a function of $\Gamma$. Each point is the median of $10$ different runs, with the error bars corresponding to the median absolute deviation about the median. The red dash line represents a hypothetic step function which would be obtained for $T_{W}\rightarrow\infty$. The two different regimes defined by the step function are named by roman numbers I and II. The value given for $\Gamma_{c}$ is tentative}
\label{jamming}
\end{figure}

On Fig.\ref{jamming}(a) the values of $\Phi$ as a function of $T_{W}$ are displayed on a normal-log representation. It can be seen that for high acceleration, a steady state is reached after a short transient, indicating the existence of a well defined mean time spent in the jammed state. However, for intermediate accelerations ($\Gamma=0.042,~0.029$ and $0.017$), no steady state is reached and the definition of an average time for the jams does not make sense. Finally, $\Phi \rightarrow 1$ for $\Gamma \rightarrow 0$ because after a finite time the flow is blocked by an arch at the outlet, which is not likely to be destroyed in the time scales involved in the experiment. Note that the slow dynamics (the time is represented in a log scale) have some features similar to aging processes. An alternative way to describe this transition is shown on Fig.~\ref{jamming} (b), where the values of $\Phi$ at $T_{W}=1000~s$ are represented as function of $\Gamma$ for $D=2.42d$. For the statistical description, we plot the median (which is a more robust locator in the case of anomalous statistics) of 10 runs, and the error bars correspond to the median absolute deviation about the median, or MAD. As is obvious, the transition is not only evidenced by the values of $\Phi$ going from one to zero in the region $0.01<\Gamma<0.06$, but also by the large error bars, as expected.

Now, let us get back to Fig.~\ref{jamming} (a) and in particular to the data corresponding to intermediate accelerations ($\Gamma=0.042,~0.029$ and $0.017$). In those cases, no steady state is reached and $\Phi$ presents a positive slope. Therefore, it is reasonable to suppose that $\Phi$ will tend to $1$ for $T_{W} \rightarrow \infty$. Assuming this supposition, $\Phi_{\infty}(\Gamma)$ would be a step function with two regimes separated by a critical acceleration $\Gamma_{c}$ (see Fig.~\ref{jamming} (b)). For $\Gamma>\Gamma_{c}$, the mean time spent in the jammed state would be well defined, and a steady flow rate would be reached. However, for $\Gamma<\Gamma_{c}$, we could be able to get the silo flowing, but the outflow rate would not be well defined, just because the mean time spent in the jammed state would not be well defined. The existence of a $\Gamma_{c}$ is in agreement with the transition to an exponent lower than 2 (absolute value) in the power laws found for the PDF of $T_{J}$ (Fig.~\ref{histtj}).

\section{Conclusions}

We presented an experimental study of the discharge of a granular hopper with an external vibration applied at the outlet. The design allows a continuous control of the vibrational energy injected in the granular packing (via a piezoelectric transducer) as well as a continuous variation the orifice size. We were thus able to measure precisely the size $D_{min}(\Gamma)$ below which the flow is halted and its dependence with the vibration rms acceleration $\Gamma$. In the absence of vibration, the grains stop flowing below an orifice size $D_{min}(\Gamma=0)$. We show that $D_{min}(\Gamma)$ decreases with an increasing vibration and this, down to one grain size, value reached for a finite vibration amplitude. This would allow, in principle, a conceptual design suited to deliver individual grains from a hopper. This surprising feature poses many questions on the fundamental mechanism at work for blockage; at the moment, we have no clear theoretical vision or even a scaling argument to explain the relationship between $D_{min}$ and $\Gamma$. We intend to explore this question in future works by changing the frequency of vibration, the grain size and the hopper geometry.

Below the $D_{min}(\Gamma=0)$ threshold, we have evidenced an intermittent regime. It consists of flowing and jammed periods. We describe the flowing intervals as a stochastic process for the flow rate. Vibrations only affect probabilities to observe low flow rates but hardly change the distribution around the mean value set only by the hopper aperture. In a future work we will seek to provide a quantitative description of the flow statistics and its relation with the stopping probability. The jammed intervals are associated with the blockages of the packing, that can last a very long time before a continuous flow may restart. At low acceleration, we evidence an anomalous statistics for the jamming times, characterized by a power law distribution with no first moment. More precisely, we characterized the transition to an anomalous statistics by monitoring the fraction of time for which the hopper is jammed as a function of the observation time $T_{W}$. This anomalous statistics is associated with an elusive value for the blockage thresholds that may actually depend either on the total number of grains available or the time we are willing to wait for the flow to resume. This property is strongly reminiscent of the anomalous dynamics usually observed for creeping flows of glassy materials. It would be interesting to relate the flow anomalies described here to the new proposition that has emerged recently to describe glassy systems and the contribution of the so called ``soft-modes'' dynamics to transport properties.

\acknowledgments
We thank I. Zuriguel and L. A. Pugnaloni for their suggestions. We acknowledge the ``Picasso'' bilateral collaboration program between Spain and France. Funding for this work was provided by Projects FIS2005-03881, FIS2008-06034-C02-01 and HF 2006-0234 (Spanish Government). AJ thanks Fundaci\'on Ram\'on Areces for a grant.

\end{document}